# Deep generative model conditioned by phase picks for synthesizing labeled seismic waveforms with limited data *[1]

Guoyi Chen[1,2], Junlun Li*[1,2,3] and Hao Guo[4]

[1]Laboratory of Seismology and Physics of the Earth's Interior, School of Earth and Space Sciences, University of Science and Technology of China, Hefei, China, 230026

[2]Mengcheng National Geophysical Observatory, University of Science and Technology of China, Hefei, China, 230026

[3]CAS Center for Excellence in Comparative Planetology, Hefei, China, 230026.

[4]Department of Geoscience, University of Wisconsin-Madison, Madison, WI, USA, 53706

*Corresponding author: lijunlun@ustc.edu.cn

---

# Abstract

Shortage of labeled seismic field data poses a significant challenge for deep-learning related applications in seismology. One approach to mitigate this issue is to use synthetic waveforms as a complement to field data. However, traditional physics-driven methods for synthesizing data are computationally expensive and often fail to capture complex features in real seismic waveforms. In this study, we develop a deep-learning-based generative model, PhaseGen, for synthesizing realistic seismic waveforms dictated by provided P- and S-wave arrival labels. Contrary to previous generative models which require a large amount of data for training, the proposed model can be trained with only 100 seismic events recorded by a single seismic station. The fidelity, diversity and alignment for waveforms synthesized by PhaseGen with diverse P- and S-wave arrival labels are quantitatively evaluated. Also, PhaseGen is used to augment a labelled seismic dataset used for training a deep neural network for the phase picking task, and it is found that the picking capability trained with the augmented dataset is unambiguously improved. It is expected that PhaseGen can offer a valuable alternative for synthesizing realistic waveforms and provide a promising solution for the lack of labeled seismic data.

# 1. Introduction

Seismic waveform synthesis can generate theoretical seismograms and has been used widely to interpret observed seismic data (Mousavi and Beroza, 2023), or invert for the subsurface structures (Tromp, 2020; Li et al., 2023). In general, waveform synthesis relies on the elastodynamic equations which are related to the specific properties of seismic sources and the subsurface structures to infer the waveforms observed at seismic stations. Common numerical methods for synthesizing seismic waves include the finite difference (e.g., Moczo *et al.*, 2007) and the spectral element methods (e.g., Komatitsch and Tromp, 1999), which are effective in solving a variety of problems in seismic wave propagations. Still, these numerical methods have limitations. Not only these methods often require significant computational resources to synthesize seismic waveforms, but also they may struggle to accurately capture detailed features in the high-frequency waveforms, since the complex structures of the real Earth is rarely known (Florez *et al.*, 2022).

Recently, simulations of realistic waveforms have become more imperative with the development of deep-learning (DL) seismology (Mousavi and Beroza, 2022) when researchers attempt to train DL models through synthetic waveforms due to lack of real data (Mousavi and Beroza, 2023). However, due to the discrepancy between real and synthetic seismic waveforms, it is often challenging to directly apply deep neural networks trained by synthetic dataset to process field data (Alkhalifah *et al.*, 2021; Birnie and Alkhalifah, 2022).

Instead, generative models (Jebara, 2012) which learn the statistical distribution of real seismic data and then sample from the learned distribution can be used as a potential alternative to the physics-based seismic wave simulations. Since the advent of deep learning, a variety of methods referred to as the deep generative models (DGMs) have emerged by combining generative models with deep neural networks, such as the variational autoencoder (VAE, Kingma and Welling, 2013), generative adversarial network (GAN, Goodfellow *et al.*, 2014) and the diffusion model (Ho *et al.*, 2020). DGMs possess the capability to map a simple distribution (e.g., Gaussian) to a more complex distribution such as observed seismic data (Ruthotto and Haber, 2021). With the rapid advancement in GPU computing, thousands of synthetic waveforms can now be conveniently generated within a matter of seconds. However, synthesizing waveforms without any prior constraint is of limited value in practice, since such waveforms cannot reflect the characteristics of the observation systems, subsurface structures, or seismic sources. Instead, waveforms constrained by prior knowledge can be synthesized using a specific DGM called conditional generative adversarial network (cGAN). By serving prior rules into cGAN as conditions, cGAN can generate diverse waveforms which comply with specified rules. Wang *et al.* (2021) proposed a cGAN-based DGM called SeismoGen to synthesize waveforms from binary conditions, i.e., the waveform either contains an earthquake or pure noise. In their study, SeismoGen was first trained by 6,150 seismic records from a seismic station in Oklahoma, and then used to augment a dataset for training a neural network designed for earthquake detections, which achieved better performance compared to its counterpart trained only with the original un-augmented dataset. Novoselov *et al.* (2021) further expanded the input conditions of cGAN from simple binary classifications to more complicated ones such as epicentral distances and earthquake magnitudes. By learning millions of waveforms from the STanford EArthquake Dataset (STEAD, Mousavi *et al.*, 2019), their cGAN model was able to generate waveforms complying with various labels from different categories. Florez *et al.* (2022) successfully synthesized accelerograms conditioned by magnitudes, epicentral distances and the time-averaged shear-wave velocity measured over the top 30 m ($V_{S30}$) in the subsurface. Their cGAN was trained using 106,189 seismic traces from the Japanese seismograph networks K-NET and KiK-net. Using a similar dataset for training, Esfahani *et al.* (2023) synthesized amplitude spectra of ground-motion recordings in the time-frequency domain using cGAN in an attempt to simplify waveform synthesis, and then reconstructed corresponding phase information from amplitude spectra with a novel phase retrieval algorithm. While these studies require a large number of seismic recordings to train a generative model, the number of training samples and labels, however, can often be insufficient in the fields of geosciences (Li *et al.*, 2023). In addition, it is difficult to quantify whether the synthesized waveforms comply with the prior conditions in the previous studies.

In this study, we develop a new cGAN model named PhaseGen which can be trained with a small dataset to generate realistic seismic waveforms constrained by prior P- and S-wave arrivals. The generation ability of PhaseGen, in terms of fidelity, diversity and alignment with conditions, is quantitively evaluated in this paper. Also, by comparing the performances of a representative DL-based phase picker trained with datasets before and after augmentation, the feasibility to train DL models by augmenting small labeled dataset with realistic synthetic waveforms is also proved.

## 2. Methodology

### 2.1 Background of conditional GAN

The generative adversarial network (GAN) is a deep learning framework that trains two neural networks, a generator and a discriminator, in an adversarial manner (Goodfellow *et al.*, 2014). The generator produces synthetic data and the discriminator evaluates its authenticity, leading to the creation of new data samples that closely resemble the original training data. In comparison, the conditional GAN (cGAN) is an extension of the standard GAN framework by incorporating external input information which can tailor generated samples to specific characteristics (Mirza and Osindero, 2014). Specifically, an objective function guided by the discriminator is constructed to minimize the discrepancy between the true and generated data distributions. Contrary to the original GAN which applies the Jensen-Shannon divergence as the measurement for the discrepancy (Goodfellow *et al.*, 2014), in this study we adopt a more sophisticated objective function $L$ which minimizes the Wasserstein distance between the two distributions (Gulrajani et al., 2017; Zheng et al., 2020):

$$L = L_W + \lambda L_{GP}, \qquad (1)$$

where $L_W$ is proportional to the Wasserstein distance, $L_{GP}$ is a gradient penalty term satisfying the Lipschitz constraint of the Wasserstein distance-based model (Gulrajani *et al.*, 2017), and $\lambda$ is the weight for $L_{GP}$, which is set to one in our study. $L_W$ can be further expressed as:

$$L_W = \mathbb{E}_{x \sim \mathbb{P}_r}[D(x|y)] - \mathbb{E}_{z \sim \mathcal{N}(0,1)}[D(G(z|y)|y)], \qquad (2)$$

where $\mathbb{P}_r$ is the distribution of the real data samples in the training dataset, $x$ is the real data sampled from $\mathbb{P}_r$, $z$ is the random vector sampled from the standard Gaussian distribution, $y$ is the condition to control the generated samples; $G(\cdot)$ and $D(\cdot)$ represent the generator and discriminator, respectively. The gradient penalty term $L_{GP}$ is:

$$L_{GP} = - \mathbb{E}_{\hat{x} \sim \mathbb{P}_{\hat{x}}}[(\|\nabla_{\hat{x}} D(\hat{x}|y)\|_2 - 1)^2], \tag{3}$$

where $\nabla$ is a differential operator, $\hat{x}$ is sampled from a mixed distribution of real and generated data $\mathbb{P}_{\hat{x}}$:

$$\hat{x} = \varepsilon x + (1 - \varepsilon) G(z|y), \tag{4}$$

where $\varepsilon$ is sampled from a uniform distribution ranging from 0 to 1. The training procedure follows an adversarial minmax game between the generator and discriminator by optimizing the loss function:

$$\min_G \max_D L. \tag{5}$$

In the training procedure, the discriminator $D$ and generator $G$ are optimized alternately: the parameters of the discriminator remain fixed while the generator is trained, and vice versa. The loss function $L$ is further decomposed into $L_d$ and $L_g$, which represent the loss of the discriminator and generator, respectively. Taking into account Equations (1) through (5), the optimization of the discriminator can be expressed as:

$$\max_D L_d = \mathbb{E}_{x \sim \mathbb{P}_r}[D(x|y)] - \mathbb{E}_{z \sim \mathcal{N}(0,1)}[D(G(z|y)|y)] \\ - \lambda \mathbb{E}_{\hat{x} \sim \mathbb{P}_{\hat{x}}}[(\|\nabla_{\hat{x}} D(\hat{x}|y)\|_2 - 1)^2], \tag{6}$$

and the optimization of the generator can be expressed as:

$$\min_G L_g = -\mathbb{E}_{z\sim\mathcal{N}(0,1)}[D(G(z|y)|y)]. \qquad (7)$$

Through the alternative optimizations for $L_d$ and $L_g$, the adversarial minmax game in Equation (5) is realized.

In this study, PhaseGen is designed and trained based on the framework of cGAN to generate waveforms constrained by given P- and S-wave arrivals. In PhaseGen, $x$ is the real waveform in the training dataset. Since the waveform features vary with the differential times between P- and S-wave arrivals rather than the absolute arrival times, the condition $y$ in PhaseGen is simplified from respective P- and S-wave arrivals to the differential times $t_{S-P}$. Figure 1 shows the training procedure of PhaseGen, where fake waveforms are synthesized by the generator (G in Fig. 1), and the discriminator (D in Fig. 1) discerns the difference between fake (synthetic) and real waveforms.

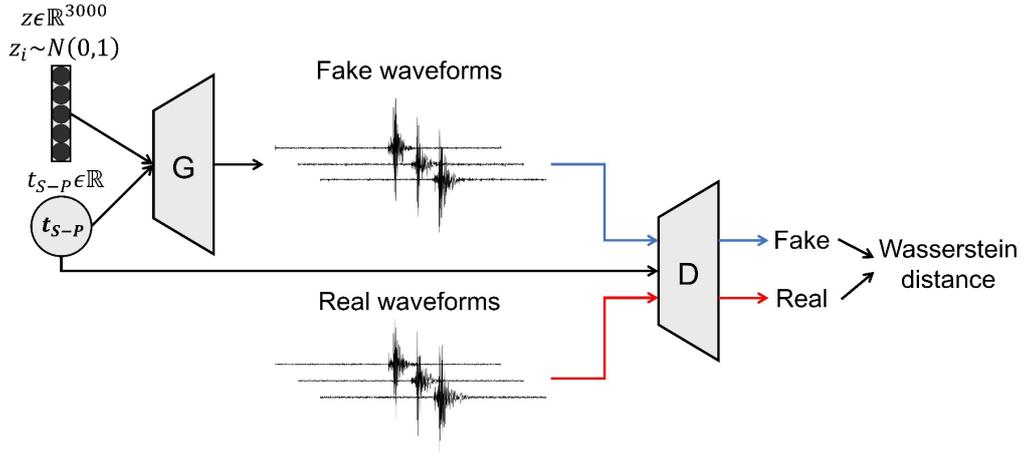

**Figure 1.** Overview of the training procedure for PhaseGen. The generator and discriminator are denoted by 'G' and 'D', respectively. $t_{S-P}$ denotes the differential time between the P- and S-

wave arrival times in the real waveforms, and is the condition ($y$) for PhaseGen.

## 2.2 Schematic workflow of PhaseGen to generate conditioned waveforms

Figure 2a illustrates the entire workflow of PhaseGen to synthesize waveforms with specified P-wave arrival times ($t_P$) and S-wave arrival times ($t_S$), which are first simplified to the differential time $t_{S-P}$, and the position of the P-arrival is fixed at the center ($t_c$) of the output waveform segment. The length of the generated waveforms is therefore twice the length of the target waveforms. Subsequently, a time window of which the starting point is set at $t_c - t_p$ is used to trim the generated segment into the target waveform. Since only one condition is applied to constrain the synthetic waveform in this way, training of PhaseGen becomes more stable. Figure 2b shows how the synthetic waveforms vary with the input $t_{S-P}$ conditions while the latent variable $z$ is fixed, whereas Figure 2c shows how the synthetic waveforms vary with the latent variable $z$ under the same input condition $t_{S-P}$.

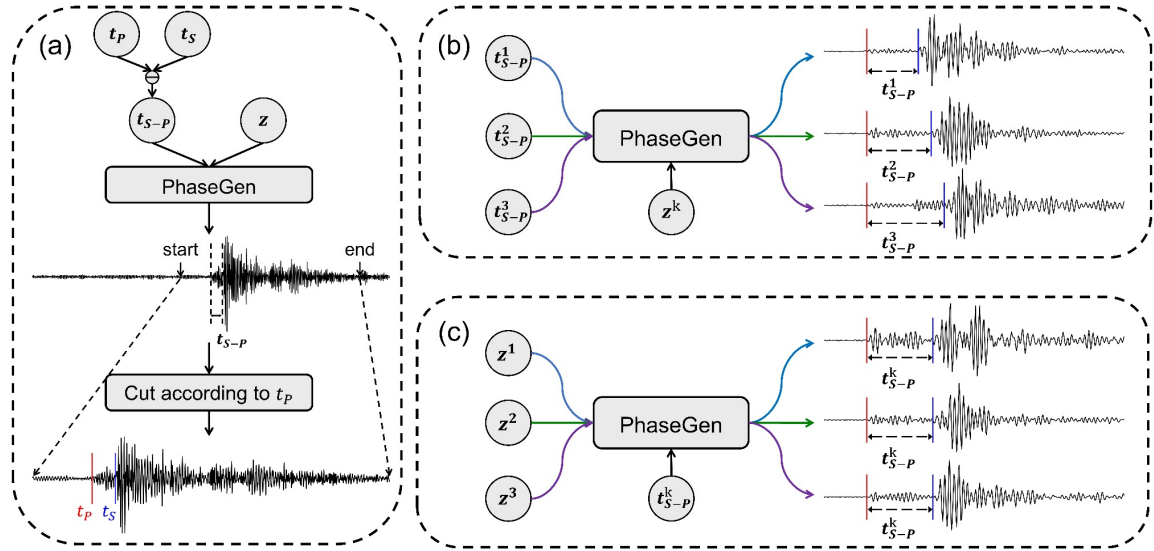

**Figure 2.** Schematic illustrations for waveform synthesis with PhaseGen. (a) Entire workflow of PhaseGen to synthesize a waveform with specified P- and S-arrival labels. (b) Illustration of how the synthetic waveforms vary with the input $t_{S-P}$ conditions while the latent variable $z$ is fixed. (c) Illustration of how the synthetic waveforms vary with the latent variable $z$ while $t_{S-P}$ is fixed.

## 2.3 Network design details for PhaseGen

The network architectures of the generator and discriminator in PhaseGen are shown in Figure 3. For the generator, the two input components which include a vector $z$ in a standard Gaussian distribution with a length of 3,000 points and the differential time $t_{S-P}$ are initially encoded with several linear or convolutional layers (Fig. 3a). Subsequently, the encoded parts are merged along the channel dimension and passed through three distinct blocks (Fig. 3b) to generate diverse waveforms for three different components, respectively. Finally, the outputs of these blocks are concatenated along the channel dimension, forming the synthetic three-component seismic waveforms. Given that the duration of the synthesized waveform is 60 s and the sampling frequency of the training dataset to be discussed later is 50 Hz, each of the output waveforms comprises 3,000 sampling points. The architecture of the ResBlock that mainly consists of six residual blocks is shown in Figure 3b. Residual connections (He *et al.*, 2016) are used in each block to facilitate direct information transmission between shallow and deep layers.

For the discriminator in PhaseGen, a deep convolutional network is used (Fig. 3c). Similar with the generator, the input waveform $x$ and $t_{S-P}$ are initially encoded separately and then merged together. Three convolutional layers are then applied to extract data features, followed by global max pooling to compress the data length and extract key features. Then 1x1 convolutional layers are used to transform channel-wise features and reduce the channel dimension to one, and the output of the discriminator is denoted as '$s$' in Figure 2d. LeakyReLU (Mass *et al.*, 2013) is chosen as the activation function for both the generator and discriminator, which introduces minor nonlinearity from negative input values to enhance the ability of the network in modeling complex patterns (Radford *et al.*, 2015). Instance normalization (Ulyanov *et al.*, 2016) is also implemented to facilitate convergence. The parameters of both networks are initialized by normal distribution with a mean of 0 and a standard deviation of 0.02, and optimized with the Adam optimizer (Kingma and Ba, 2014). We adopt the 'two time-scale update rule' (Heusel *et al.*, 2017) as our training strategy, which sets a larger learning rate for the discriminator (0.0004) than for the generator (0.0001) to achieve a better approximation of Wasserstein distance. Both the generator and discriminator are trained with 10,000 epochs, with a batch size of 64 and shuffled data orders.

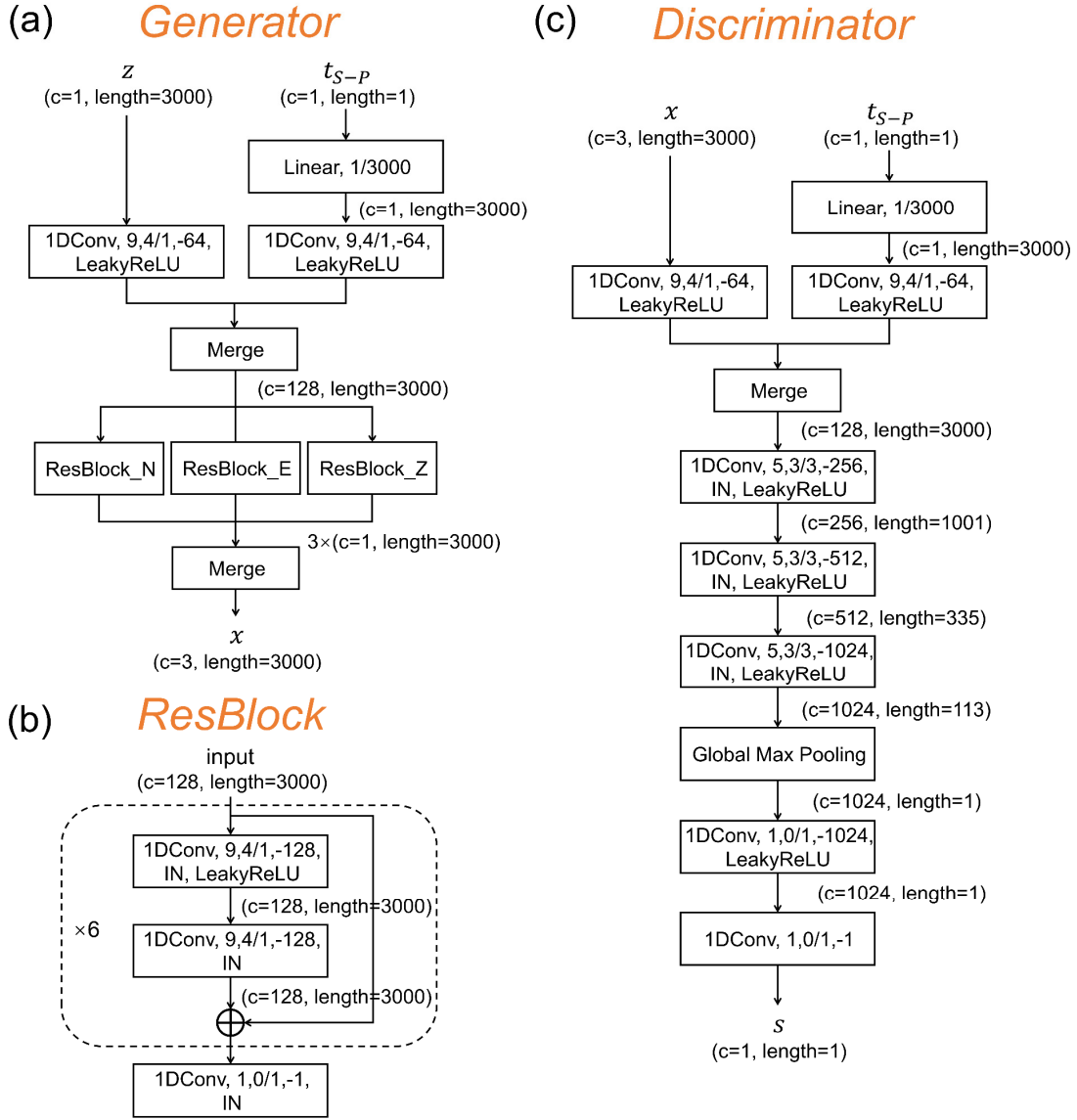

**Figure 3.** Network designs for PhaseGen. (a) Structure of the generator. (b) Detailed structure of the ResBlock (ResBlock_N, ResBlock_E, and ResBolck_Z in (a)) for generating waveforms in the three components. The dashed box indicates the residual block, with 'x6' indicating six

repetitions of the same block. (c) Structure of the discriminator. In all subfigures, 'Linear' denotes a linear transformation layer; '1/3000' indicates the dimensions for the input and output are 1 and 3,000, respectively; '1DConv' represents a 1D convolutional layer, and '1DConv, 9, 4/1, -128, IN, LeakyReLU' for instance, represents a 1D convolutional layer with a kernel size of 9, padding size of 4, stride of 1, and 128 output channels, instance normalization, and the LeakyReLU activation function following the convolutional layer; 'Merge' refers to concatenation of different inputs along the channel dimension. The number of output channels and length of each operation are indicated in the brackets in the subfigures.

## 3. Datasets and preprocessing

To evaluate the generation quality, PhaseGen is applied to an earthquake dataset acquired by 16 three-component ocean bottom seismometers (OBSs) in 2008 around the Gofar transform fault in the East Pacific Rise (McGuire et al., 2012). The P- and S-wave arrivals was initially picked using the short-term average to long-term average ratio method (Stevenson, 1976; McGuire et al., 2012). Subsequently, Guo *et al.* (2018) used a repicking method based on Akaike information criterion (Maeda, 1985; Zhang *et al.*, 2003) to further enhance the quality of the S-arrival labels. We select 4,452 events from the catalog of Guo et al. (2018) with both P- and S-wave arrivals picked at the station Gp06 (Fig. 4a). We further select 100 events randomly from these 4,452 events (red dots in Fig. 4a) to construct the training dataset for PhaseGen, and the distribution of $t_{S-P}$ with earthquake magnitudes is shown in Figure 4b. The size of the training dataset is deliberately retained small to demonstrate that the PhaseGen can be trained by a limited number of data samples.

Additionally, another 200 events (yellow dots in Fig. 4a) are chosen randomly and divided equally into two groups for training two alternative PhaseGen models, which are used to verify that generation quality is stable even with different training datasets of a limited size. The remaining 4,152 events (green dots in Figure 4a) constitute the validation dataset (100 events) and one of the test datasets (baseline A, 4052 events), which are used to evaluate the phase-picking performance of a DL-based phase picker trained with and without the augmented waveform data generated by PhaseGen. The other test dataset, baseline B, is constructed with 819 events with P- and S-wave arrival labels from the station Gp02 and 164 events from Gp16 (Fig. 4a) to further evaluate the generalization of the phase picker trained with augmented datasets.

The waveform recordings are first processed in a multistage workflow before fed to train PhaseGen. The first step is to cut event waveforms from the continuous data. For the training dataset used by PhaseGen, the event waveform windows are 60-s-long with the P-wave arrivals fixed at 30 s. We then rotate the extracted three-component waveforms to the standard vertical (Z), north (N) and east (E) components. Subsequently, the waveforms are processed in several consecutive steps, including instrument response removal, demeaning, detrending, tapering, bandpass filtering and resampling. The bandpass filter is set to 5-12 Hz, consistent with that used in McGuire *et al.* (2012) and Guo *et al.* (2018). All the waveforms are resampled at 50 Hz.

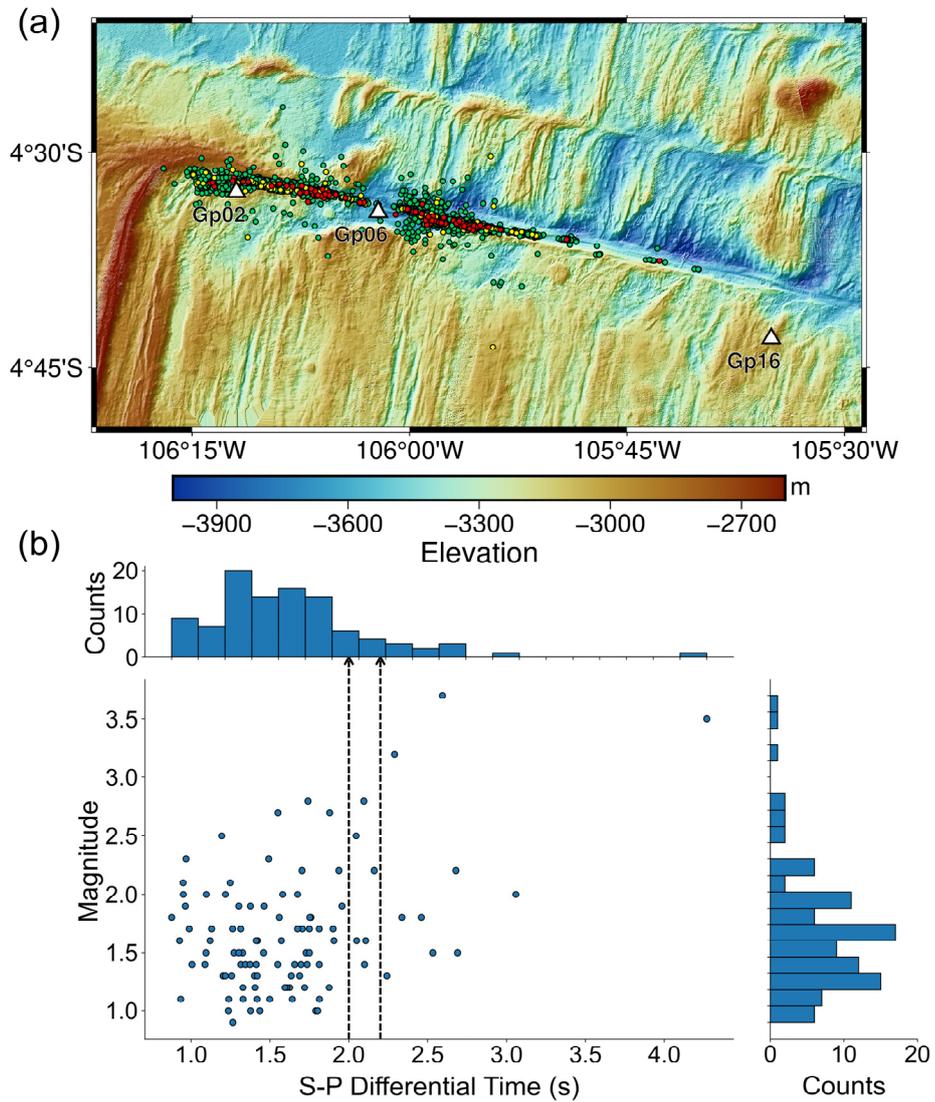

**Figure 4.** Seismicity along the Gofar transform fault in the East Pacific Rise used in this study. (a) Distribution of seismicity and the used OBS around the transform fault. The epicenters of the 100 earthquakes used for training PhaseGen are indicated by the red dots, while the yellow dots indicate the epicenters of the events used for training two alternative PhaseGen models for comparison. The green dots indicate the epicenters of the events used for testing the capability of a DL-based phase picker trained with the augmented dataset from PhaseGen. The three OBS (Gp02, Gp06,

and Gp16) discussed in this study are denoted by the white triangles. (b) Joint distribution of event magnitudes and the differential times $t_{S-P}$ for the 100 events used for training PhaseGen (red dots in Fig. 4a). The distribution histograms are shown on top and on the right. The two dashed lines mark the positions of $t_{S-P}$ equal to 2 and 2.2 s. Diversity of generated waveforms conditioned by $t_{S-P}$ within this bound is further discussed in section 4.2.

# 4. Results

## 4.1 Qualitative inspections of generated waveforms

Figure 5 shows the comparison between the real and generated three-component waveforms. It is challenging to distinguish the real and generated waveforms visually, and the input P- and S-wave arrivals match well with the onsets of the corresponding phases in the generated waveforms. Possibly due to coupling issues between the seismometer and the ocean bottom, the waveforms recorded in the horizontal N- and E-components of Gp06 exhibit more distinct low-frequency characteristics with longer oscillations. These features in the observed waveforms can be effectively mimicked by the well-trained generative models in the synthesized waveforms. The generation results of two alternative PhaseGen models trained with different events (yellow dots in Figure 4a) are shown in Figure S1. The comparison of these results indicates that changing the training sets does not notably impact the generation quality of PhaseGen.

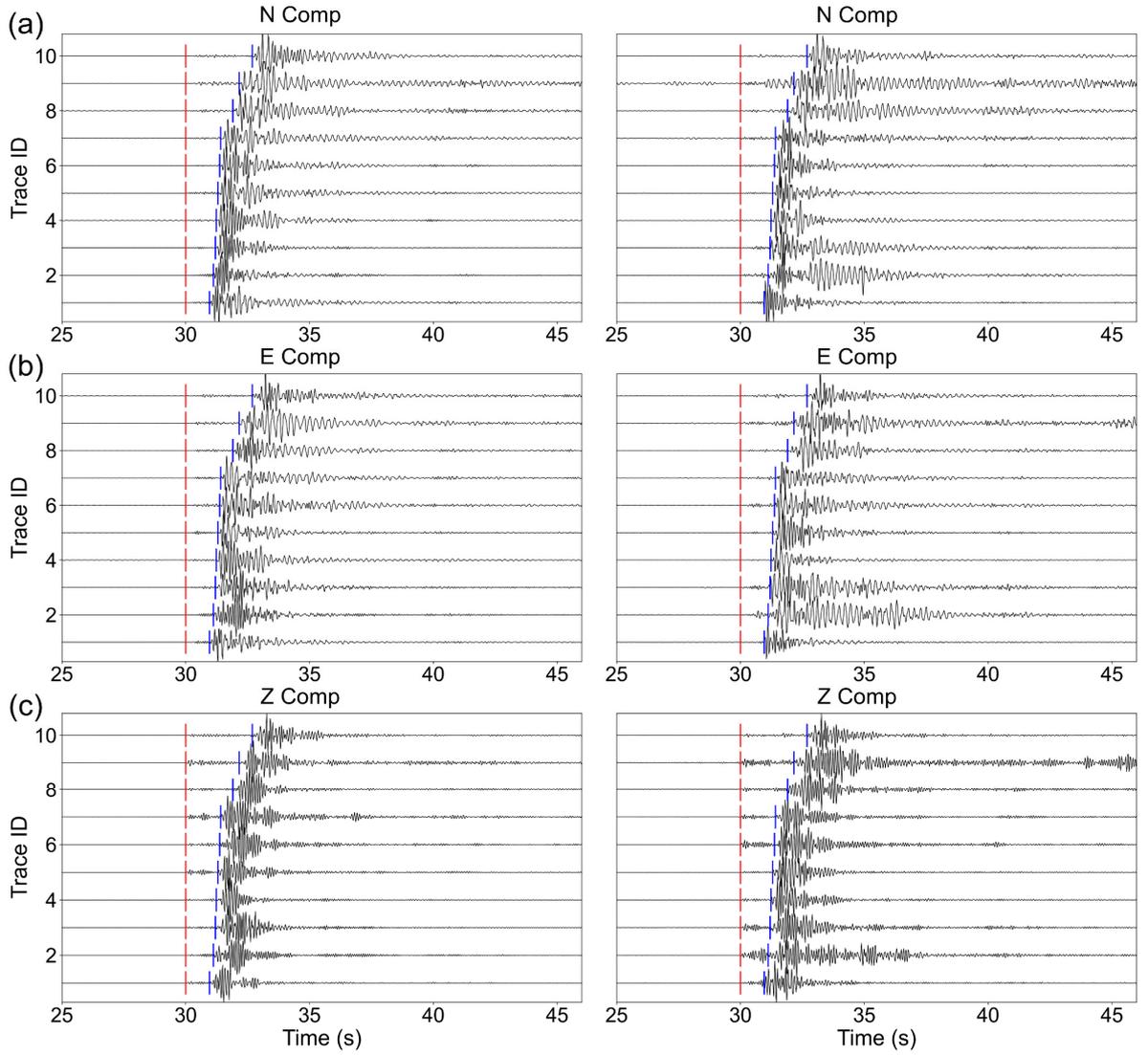

**Figure 5.** Comparison between the real and generated waveforms in three components at OBS Gp06: (a) the N-component, (b) E-component and (c) Z-component. In each subfigure, the left panel shows 10 real waveforms recorded by Gp06 with varying $t_{S-P}$, while the right panel shows 10 synthetic waveforms by PhaseGen which have identical $t_{S-P}$ with their real-waveform counterparts of the same trace ID. The real seismic waveforms are randomly selected from 10 events, and then sorted with increasing $t_{S-P}$. The red bars indicate the P-arrivals, and the blue bars

indicate the S-arrivals.

Mode collapse is a common phenomenon and remains a challenge in the training of GAN-based models (Goodfellow *et al.*, 2014), which occurs when the generator fails to capture the feature diversity in the real data and produces a limited number of samples or modes. To avoid mode collapse in this study, it is required that waveforms with diverse features can be generated even with the same $t_{S-P}$ condition, which poses a challenge for PhaseGen trained with only 100 samples. Nevertheless, Figure 6 shows 3 real data samples conditioned by different $t_{S-P}$, for each of which 4 waveform samples with diverse features are generated with varying latent variable $z$. While all the generated waveforms rigorously comply with the given $t_{S-P}$ conditions, they are still similar to the real samples apparently, indicating that PhaseGen learns varying features from different real seismic waveforms in the training dataset and avoids generating rather similar waveforms.

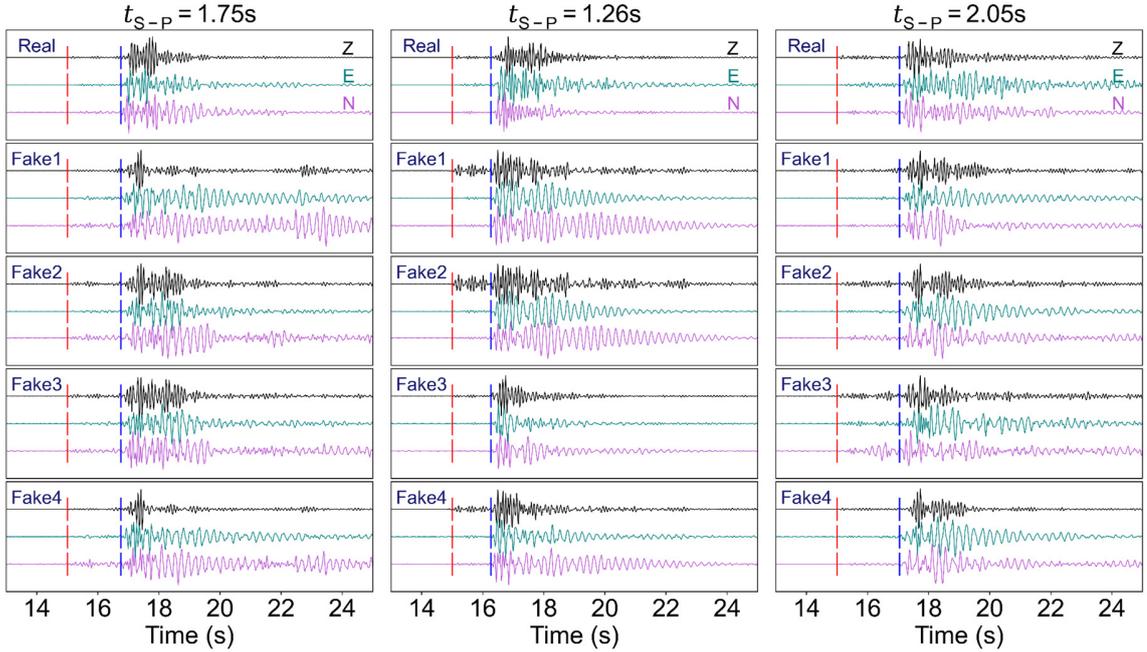

**Figure 6.** Comparisons between 3 real data samples with varying $t_{S-P}$ and the corresponding generated waveforms with the same differential time $t_{S-P}$ but 4 different latent variables $z$ sampled from standard Gaussian distributions. Each column compares real and fake waveforms of a specific $t_{S-P}$, where the first row shows the real waveforms, and the next four rows show generated waveforms based on 4 randomly sampled latent variable $z$. The N-, E-, and Z- components of the waveforms are drawn in purple, cyan, and black, respectively. The P- and S- arrivals are marked by the red and blue bars, respectively.

However, PhaseGen exhibits limited capability in synthesizing waveforms with conditions barely covered in the training dataset. In such cases, PhaseGen shows significant mode collapse. Additionally, the generator may struggle to generate waveforms satisfying conditions dissimilar with those in the training datasets. Figure 7 shows these issues using three scenarios with $t_{S-P}$ equal to 0.5 s, 3 s and 3.5 s, respectively. It should be emphasized that no training sample has labels close to these differential times (Fig. 4b). Interestingly, PhaseGen is not able to generate waveforms complying with the first condition, but it manages to generate waveforms still complying with the latter two conditions. Nevertheless, in all three scenarios mode collapse occurs, and it is suggested that for PhaseGen to generate diverse waveforms and avoid mode collapse, the training dataset should contain at least a few samples having similar labels with the input condition. In the following, the $t_{S-P}$ condition for the generator is constrained within the range 1.0 s to 2.2 s, where a comparatively dense distribution of conditions exist in the training dataset (Fig. 4b).

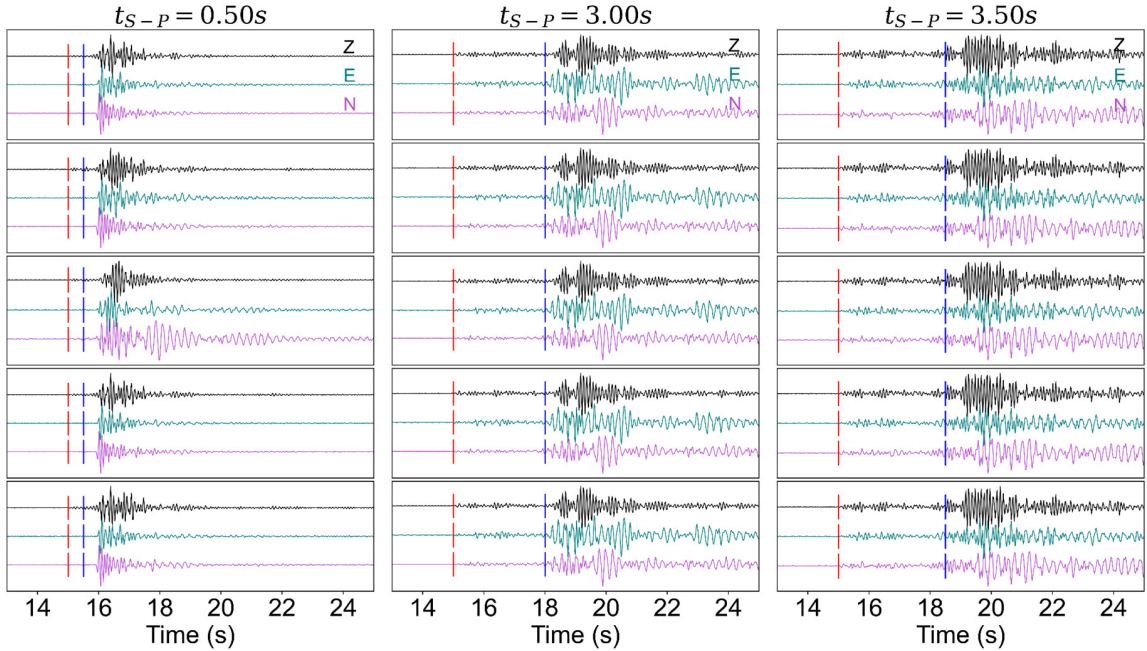

**Figure 7.** Examples of generated waveforms constrained by conditions not presented in the training dataset. Each column corresponds a different $t_{S-P}$ condition, and each row shows waveforms generated by PhaseGen with a different latent variable $z$. The rest is the same as Figure 6.

## 4.2 Quantitative analysis of generated waveforms on fidelity, diversity and alignment with conditions

After training, PhaseGen can establish a mapping from the standard Gaussian distribution to the data distribution of training dataset, thus the generator network can produce realistic yet diverse waveforms which satisfy the input conditions. The crucial questions to consider are to what extent PhaseGen imitates the training dataset, and the degree of novelty it can possess in terms of generating realistic fake data. In this section, we quantitatively analyze the generated waveforms on fidelity, diversity and alignment with given conditions.

The fidelity of PhaseGen is evaluated by measuring the similarity between the generated and real waveforms:

$$p = \frac{\sum_{i=1}^{n}(x_i - \bar{x})(x_i' - \overline{x'})}{\sqrt{\sum_{i=1}^{n}(x_i - \bar{x})^2 \cdot \sum_{i=1}^{n}(x_i' - \overline{x'})^2}}, \qquad (8)$$

where $p$ is the normalized correlation coefficient ranging from -1 to 1, $x$ and $x'$ denote the real and generated three-component waveforms, respectively, $\bar{x}$ and $\overline{x'}$ represent the corresponding means, and $n \in \{1,2,3\}$ is the component index. Strong positive or negative correlations are identified with $p$ close to 1 or -1, respectively, and weak correlation is identified with $p$ close to 0. We generated 1,000 fake samples randomly and calculated $p$ between each fake sample and all the 100 real seismic waveform samples in the training dataset for PhaseGen. Here one sample is defined as a set of three-component waveforms. Figure 8 shows the statistics of the maximum $p$ for each fake sample. The mean of the maximum $p$ is 0.84, indicating an unambiguous correlation between the generated and real waveforms. Though some fake samples are found to have remarkable similarities (maximum $p$ greater than 0.95) with real waveforms in the training dataset (Figure 9), most generated waveforms have relatively low similarities with the training dataset (Figure 10), exhibiting distinct diversity in data synthesis.

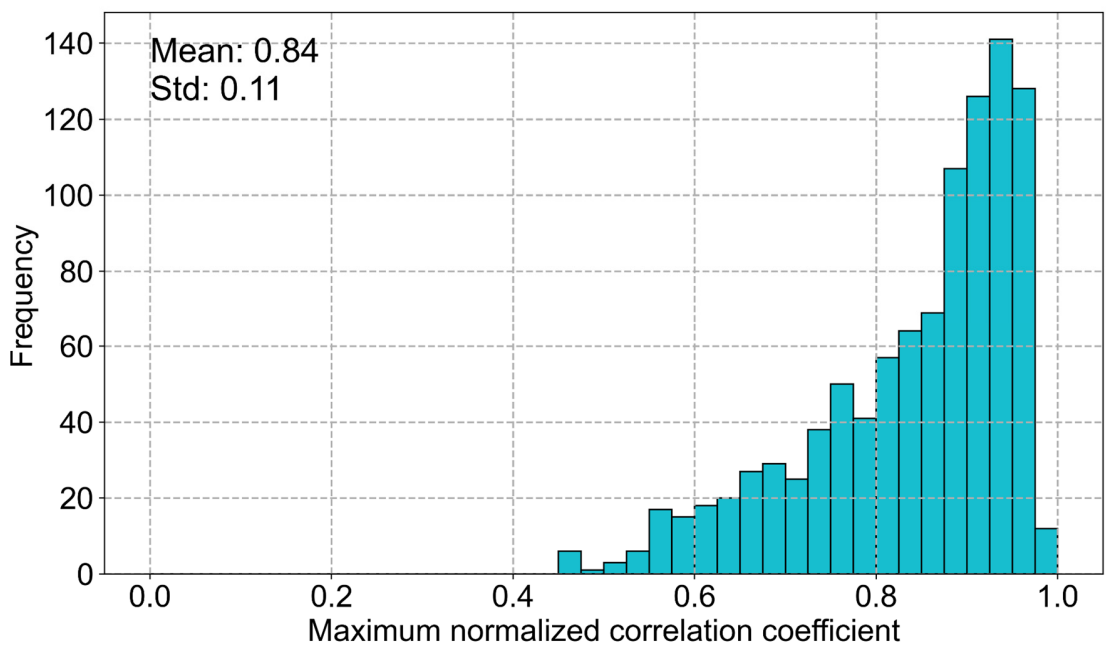

**Figure 8.** Distribution of the maximum correlation coefficients $p$ between each of the 1,000 randomly generated fake waveform samples and the 100 real waveform samples in the training dataset. The mean and standard deviation of the coefficients are indicated in the upper left corner.

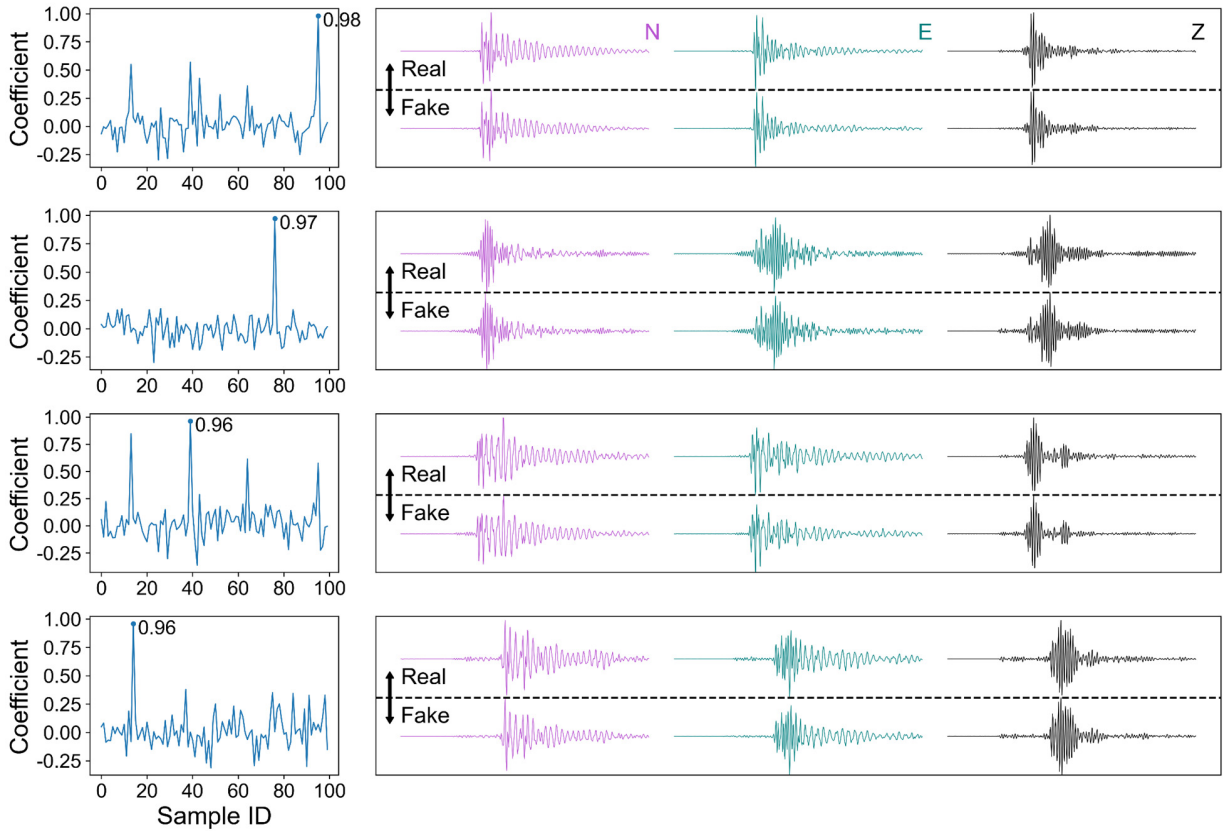

**Figure 9.** Comparisons between four generated waveform samples which exhibit high similarities with particular real waveform samples in the training dataset. The left panel in each row shows the normalized correlation coefficients $p$ between one generated waveform sample by PhaseGen and the 100 real waveforms in the training dataset, with the maximum labelled. The right panel in each row shows the generated waveform sample and the real waveform sample from the training dataset which has the highest correlation coefficient as marked in the left panel. The N-, E-, and Z-components of the waveform sample are plotted in purple, cyan, and black, respectively.

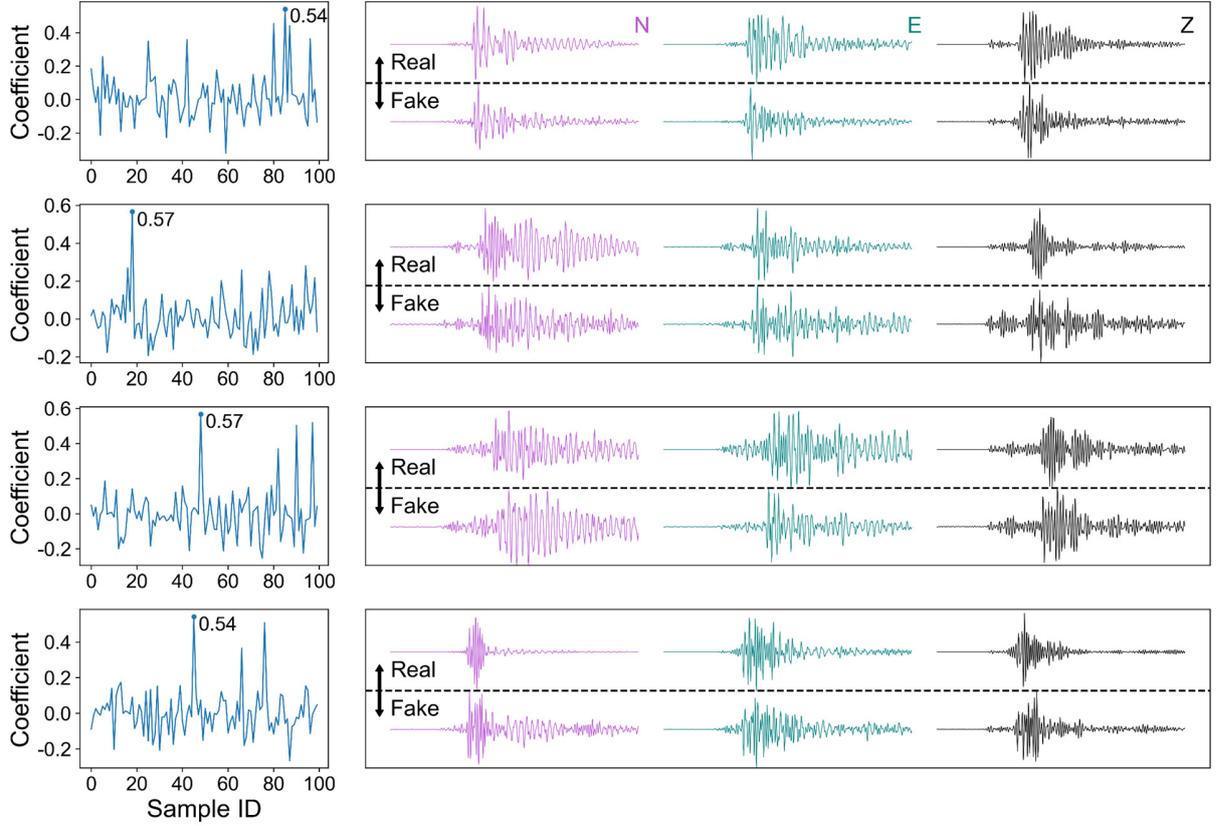

**Figure 10.** Comparisons between four generated waveform samples which exhibit relatively low similarities with the 100 real waveform sample in the training dataset and those real waveform samples which have the relatively highest $p$ with the generated ones. The rest is same as Figure 9.

To further analyze the diversity of generated waveforms, we generate a large number of waveform samples under six different conditions with $t_{S-P}$ equal to 1.0 s, 1.3 s, 1.5 s, 1.7 s, 2.0 s, and 2.2 s, respectively, and compare their similarities in each condition. Specifically, 100 fake samples are generated for each condition, and $p$ for all pairs of these 100 samples are calculated.

Figure 11 shows the correlation matrices for the calculated $p$ under the six conditions. Under the first four conditions, the generated waveforms show diversity with varying $p$. However, there is a decrease in diversity when $t_{S-P}$ increases to 2.0 s, and the diversity is further reduced when $t_{S-P}$ is increased to 2.2 s. The decreasing diversity under these two conditions can be attributed to the scarcity of waveform samples in the training dataset when $t_{S-P}$ is larger than 2.0 s (Fig. 4b) compared to the rest conditions.

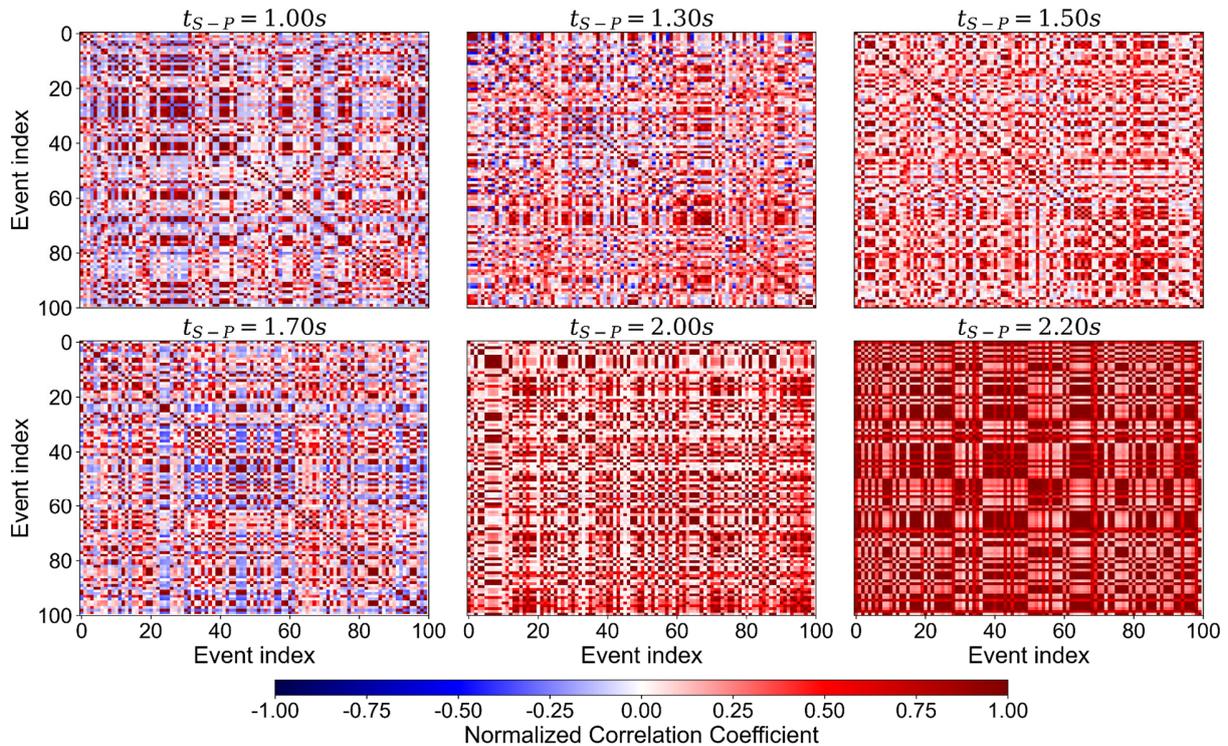

**Figure 11.** Correlation matrices for the normalized cross-correlation coefficient $p$ for all pairs of 100 fake waveform samples under each of the six different $t_{S-P}$ conditions.

To check the alignment between the generated waveforms by PhaseGen and the given conditions, that is, to verify the consistency between the phase onsets in the waveforms and the provided P- and S-wave arrival labels, we generate 1,000 waveform samples with random $t_{S-P}$ ranging from 1 s to 2.2 s, and then trim the waveforms with random $t_P$. In this way, the waveform samples are generated with entirely randomized P- and S-arrivals, which are then manually picked without any prior knowledge. Figure 12 shows the distribution of discrepancies between the manual picks and the input conditions, where positive discrepancies ($\delta P, \delta S > 0$) indicate that manual picks are later than the input conditions, and vice versa. The average discrepancy for the P-arrivals (8.6 ms) is less than the sampling interval (20 ms), while that for the S-arrivals (30.4 ms) is about 1.5 times the sampling interval, indicating the input conditions are well aligned in the synthetic waveform samples.

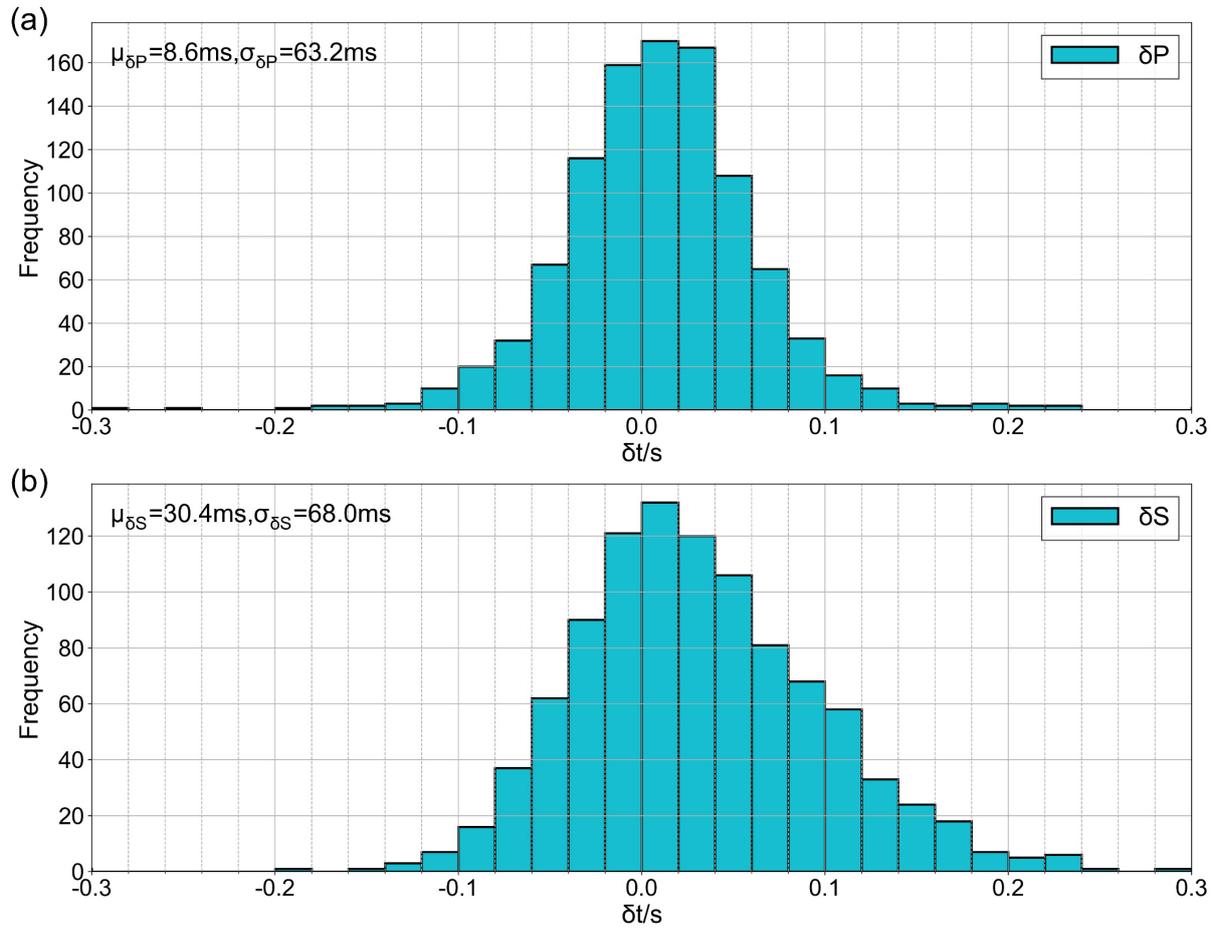

**Figure 12.** Distribution of discrepancies between the manual picks and input conditions of the generated waveforms: (a) discrepancies for the P-arrivals, (b) for the S-arrivals. The mean ($\mu$) and standard deviation ($\sigma$) of the discrepancies are marked in the upper left corners.

# 5. Discussion

## 5.1 Generation of realistic seismic traces for data augmentation

Data augmentation is a technique that generates new realistic training samples based on collected datasets, thereby increasing both the quantity and diversity of the training samples, which in turn, enhance the performance and robustness of deep learning models (Zhu *et al.*, 2020). It is often difficult to train a stable DL-based seismic arrival picker with a limited number of training data (Zhang *et al.*, 2021). In the previous sections, we demonstrate that PhaseGen can learn waveform features from a small set of samples and generate diverse waveforms with varying P- and S-arrival labels. In the following, dataset enrichment with the diversely generated samples for improving the performance of a DL-based phase picker is discussed.

In this study, PhaseNet, a representative seismic phase picking network (Zhu *et al.*, 2019) based on the U-Net architecture (Ronneberger *et al.*, 2015), is trained by the OBS data with and without augmentation, and the corresponding picking performances are compared. It should be emphasized that PhaseNet is a DL-based phase picker, while the proposed PhaseGen is a generative model to synthesize conditioned realistic waveforms, though their names may seem similar. Input with three-component waveforms from a seismic station, PhaseNet outputs the probabilities of P-arrival, S-arrival and noise for each sampling point in the waveforms. This neural network and its variations have been extensively applied in earthquake monitoring and location (e.g., Park *et al.*, 2020; Wang *et al.*, 2020; Chen and Li, 2022). We use the same 100 labelled real events (red dots in Fig. 4a) for training PhaseGen as the un-augmented training dataset to train PhaseNet from scratch.

Five groups of augmented training dataset, G100, G300, G500, G1000, and G3000 containing different numbers of generated three-component waveform samples, are also organized with additional labelled synthetic waveforms by PhaseGen. Each group contains 10 distinct sub-datasets with diverse waveforms generated by PhaseGen, which allows for a statistical evaluation of the performance of the trained PhaseNet. For instance, each sub-dataset in G300 contains 300 diversely generated waveform samples from PhaseGen and the same batch of 100 real waveform samples from the original training dataset. Various PhaseNet models are then trained with each of the augmented sub-dataset and the un-augmented dataset (100 real waveform samples only) for comparison. As mentioned in Section 3, we test the picking performance of PhaseNet with two separate benchmark datasets: 1) baseline A, which contains the real waveforms from Gp06, and 2) baseline B, which contains the real waveforms from Gp02 and Gp16. We used precision, recall, and the F1 score as three metrics to evaluate the DL-based picker:

$$Precision = \frac{T_p}{T_p + F_p}, \qquad (9)$$

$$Recall = \frac{T_p}{T_p + F_n}, \qquad (10)$$

$$F_1 score = 2\frac{Precision \times Recall}{Precision + Recall}, \qquad (11)$$

where $T_p$, $F_p$ and $F_n$ represent the counts of true positives, false positives and false negatives, respectively. In our study, a pick is deemed positive when the peak probability of the network output exceeds 0.3. When the time difference between a positive pick and a reference manual pick is less than 100 ms, it is classified as a true positive, or classified as a false positive otherwise. Valid picks misidentified as noise by the neural network are classified as false negatives. Among these three evaluation metrics, since the F1 score reflects the capabilities of phase picking and event detection of the trained network more comprehensively, it serves as a direct criterion for evaluating the performance of the phase picker.

Tables 1 and 2 show the performance details of various trained PhaseNet models using baselines A and B. Since both the un-augmented training dataset for PhaseNet and the test dataset baseline A originate from the same station (Gp06), PhaseNet performs reasonably well in picking P- and S-arrivals even without augmentation (Table 1). However, data augmentation with PhaseGen can further improve the picking performance of PhaseNet, and the degree of improvement is in general positively correlated with the number of augmented samples. When the number of augmented samples is relatively small (e.g., G100), the performance of the trained picker may even suffer from slight degradation; however, once the number of augmented samples reaches a certain threshold, the network obtains stable improvement in seismic phase picking. PhaseNet trained with G3000 shows no significant improvement compared to its counterpart trained with G1000 (Table 1), indicating that the improvement of the trained picker levels off with increasing augmented samples.

Data augmentation can also enhance the picking generalization of the trained PhaseNet. The waveforms in baseline B are obtained from two different stations, Gp02 and Gp16, which have different types of instruments (Moyer *et al.*, 2018) and ground coupling with the ocean bottom, resulting in different frequency responses in the recorded waveforms compared to the records in Gp06. Therefore, the neural network trained with the un-augmented dataset performs worse for baseline B compared to baseline A, especially for picking the S-arrivals. It is shown in Table 2 that data augmentation based on the synthetic realistic waveforms from PhaseGen effectively improves the performance of the trained phase picker for both the P- and S-wave arrivals on baseline B.

**Table 1.** Evaluation metrics for baseline A. The original dataset is from the OBS Gp06. Since each augmented dataset group contains 10 sub-datasets, the mean and standard deviation of the evaluation metrics are provided for each group.

| Training Dataset Group | Phase | Precision (%) | Recall (%) | F1 Score (%) |
|---|---|---|---|---|
| Original Dataset | P | 97.54 | 87.09 | 92.02 |
| G100 | P | 93.25±1.62 | 89.60±8.34 | 91.22±5.15 |
| G300 | P | 95.70±1.20 | 96.47±1.07 | 96.08±0.91 |
| G500 | P | 96.36±0.45 | 96.28±2.31 | 96.70±1.15 |
| G1000 | P | 96.79±0.43 | 97.84±0.88 | 97.31±0.51 |
| G3000 | P | 97.23±0.51 | 97.13±1.03 | 97.17±0.44 |
| Original Dataset | S | 95.89 | 97.69 | 96.78 |
| G100 | S | 95.06±1.10 | 95.16±4.01 | 95.07±2.44 |
| G300 | S | 96.06±0.81 | 97.92±1.11 | 96.98±0.79 |
| G500 | S | 96.70±0.46 | 98.66±0.88 | 97.67±0.55 |
| G1000 | S | 97.29±0.40 | 99.07±0.28 | 98.17±0.23 |
| G3000 | S | 97.36±0.20 | 99.09±0.32 | 98.22±0.22 |

**Table 2.** Evaluation metrics for baseline B. The original dataset is from the OBS Gp02 and Gp16. Since each augmented dataset group contains 10 sub-datasets, the mean and standard deviation of the evaluation metrics are provided for each group.

| Training Dataset Group | Phase | Precision (%) | Recall (%) | F1 Score (%) |
|---|---|---|---|---|
| Original Dataset | P | 97.61 | 88.98 | 93.09 |
| G100 | P | 93.73±1.36 | 89.43±7.50 | 91.38±4.33 |
| G300 | P | 95.17±1.25 | 96.06±1.84 | 95.58±0.86 |
| G500 | P | 96.05±1.17 | 96.10±3.30 | 96.03±1.42 |
| G1000 | P | 96.26±1.11 | 97.97±1.14 | 97.10±0.84 |
| G3000 | P | 96.43±0.80 | 97.44±1.59 | 96.92±0.72 |
| Original Dataset | S | 68.50 | 68.32 | 68.41 |
| G100 | S | 78.04±4.29 | 63.13±18.77 | 67.78±11.64 |
| G300 | S | 76.09±5.43 | 76.11±11.89 | 75.49±5.96 |
| G500 | S | 76.29±4.57 | 87.33±9.34 | 81.01±4.44 |
| G1000 | S | 78.97±4.81 | 89.84±4.49 | 83.91±3.02 |
| G3000 | S | 84.19±3.49 | 87.00±4.13 | 85.48±2.61 |

## 5.2 Critical factors for successful waveform generations

Training neural networks to synthesize seismic waveforms is understandably a challenging task since it involves construction of entire waveforms. Consequently, selection of network architectures and hyperparameters may considerably affect the quality of generated waveforms. In the following, we attempt to summarize several critical factors that influence the performance of the generator in PhaseGen.

First, proper normalization is crucial (Ulyanov *et al.*, 2016), since the normalization layer in deep neural networks accelerates the loss convergence and enhances the training stability (Ioffe *et al.*, 2015) by adjusting the mean and standard deviation of the input data. Though batch normalization is a widely used strategy (Ioffe *et al.*, 2015), we find that replacing batch normalization with instance normalization can considerably improve the generation quality. Figure S2 shows the waveforms generated by an alternative PhaseGen which is trained with batch normalization instead of instance normalization. It is clear that there is a distinct decline in the quality of generated waveforms, which may be attributed to different normalization modalities of these two techniques. While batch normalization rescales features in the entire mini-batch, instance normalization is independent of the batch size and normalizes each data sample in the mini-batch separately. Therefore, in contrast to batch normalization, instance normalization yields less dependence among individual waveforms in the generation process, which contributes to the diversity in the synthetic waveforms.

Second, the training procedure is terminated after 10,000 epochs in this study, at which $L_W$ proportional to the Wasserstein distance (Eq. 1) did not reach the minimum. Figure S3 shows that $L_W$ approaches zero after 10,000 epochs and its change gradually levels off. However, the generated waveforms by PhaseGen can lose diversity and starts to closely resemble the training waveforms when $L_W$ further decreases. Thus, the training process is terminated prematurely at 10,000 epochs to increase the diversity of generated waveforms. In comparison with Figure 8, Figure S4 shows the statistics for the maximum normalized correlation coefficients for 1,000 samples generated by an alternative PhaseGen trained with the same dataset in 20,000 epochs. Compared to PhaseGen trained with 10,000 epochs, the mean of the maximum correlation coefficients increases from 0.84 to 0.91, indicating that the synthetic waveforms become increasingly similar with the training datasets when the generative model is trained in more epochs. In fact, the mean of the maximum correlation coefficients monotonically increases with the number of training epochs, as shown in Figure S5. Thus, by adopting an early termination strategy on the training process before $L_W$ reaches zero, a balance of diversity versus similarity between the synthetic and real waveforms in the training dataset can be obtained.

Third, simplification of input conditions is instrumental in ensuring generation of high-quality waveforms with satisfactory alignments using a limited number of data samples. By aligning the P-arrival times of the waveforms in the training dataset at 30 s and using the differential time $t_{S-P}$ as the only input condition, the task complexity was effectively reduced. We also attempt to train a PhaseGen model directly using the P- and S-arrival times as two input conditions. However, the synthetic waveforms using this generative model look unrealistic (Fig. S6).

## 5.3 Potential uses for conditional data generation in seismology

Using data augmentation for training a DL-based phase picker as an example, we present a practical application of conditional seismic waveform generation. It is expected that conditional generative models should have considerable potential for broad applications in seismology. While this study focuses on synthesizing waveforms conditioned by P- and S-wave arrivals, other constraints for seismic data generations can also be incorporated. In the area of waveform synthesis, Florez *et al.* (2022) and Esfahani *et al.* (2023) studied generation of ground motion recordings constrained by earthquake magnitudes, epicentral distances and the average shear-wave velocities at different sites. The generated waveforms in their studies are consistent with observed data qualitatively and recover some features in the high-frequency wavefields, which are normally challenging for physics-based approaches. Depending on specific requirements, other conditions could also be imposed to constrain generated waveforms, such as the signal-to-noise ratio, types of earthquakes (e.g., tectonic, induced, explosion, collapse, sonic boom, etc.), as well as source mechanisms (Li *et al.*, 2011; Li *et al.*, 2023).

It should also be noticed that waveform generation by PhaseGen is stochastic. With the latent variable $z$, diverse waveforms differing from the training waveforms with the same $t_{S-P}$ condition can be generated. Since the scarcity of labeled seismic data often poses a significant challenge in utilizing deep neural networks to solve geophysical problems (Li *et al.*, 2023; Wu *et* *al.*, 2023), it is believed that PhaseGen can provide an effective approach to enriching labelled datasets, especially those that are often difficult to obtain sufficiently, such as seismograms from ocean-bottom seismometers, large-magnitude earthquakes (Hu *et al.*, 2022) and marsquakes (Lognonné *et al.*, 2023).

Furthermore, well trained DGMs can generate samples constrained by certain conditions not present in the training data. For example, by using the instrumental coordinates as conditions, the characteristics of waveforms recorded by seismometers at various locations in an array can be learned by neural networks, and seismic data hypothetically observed at locations in between can be inferred. In comparison with conventional interpolation methods for seismic data, such as the rank reduction method (i.e., Gao *et al.*, 2013) and sparse transformation method (i.e., Chen *et al.*, 2019), the DL-based conditional generation shows more flexibility since it is independent of any specific domain transformation. Though PhaseGen trained with only 100 events exhibits limited predictive capabilities for $t_{S-P}$ not present in the training dataset in this study, it is expected that the generation quality on waveforms with non-existent conditions can be further improved with more training samples (Florez *et al.*, 2022).

## 6. Conclusion

In this study, we propose a novel generative model, PhaseGen, to synthesize realistic waveforms conditioned by continuously varying P- and S-wave arrival labels using limited amount of training data recorded by an ocean bottom seismometer. PhaseGen avoids mode collapse and can generate diverse waveforms differing from those in the training dataset. We analyze the generated waveforms for their fidelity, diversity, and alignment with input conditions, and also discuss critical factors for successful training. Using data augmentation on the training dataset for a DL-based seismic phase picker as an example, we show a potential use of PhaseGen in enriching seismic waveforms that are scarce and difficult to obtain. It is expected that this data-driven DL-based realistic synthesis approach will be a valuable complement to conventional model-based numerical modeling, and its applications in various scenarios deserve more investigations in the future.

# Data and Resources

Our source code can be accessed at https://github.com/Billy-Chen0327/PhaseGen. The Gofar waveform data used in our paper can be accessed at the Incorporated Research Institutions for Seismology (IRIS) data center under the network codes ZD (https://www.fdsn.org/networks/detail/ZD_2007/). Figure 3 in this paper is plotted using the Generic Mapping Tools (Wessel *et al.*, 2019). The open-source machine learning package Pytorch (Paszke *et al.*, 2019) is used to build the deep learning models. Data are preprocessed through the Obspy package (Beyreuther *et al.*, 2010).

# Acknowledgements

This research is supported by the National Science Foundation of China under grants U2139204. GPUs used in this study for training the neural networks are provided by the Hefei Advanced Computing Center on the HG architecture. The authors thank Prof. Haijiang Zhang for his helpful suggestions in conceptualizing this paper.

eabm4470.

Mousavi, S. M., & Beroza, G. C. (2023). Machine Learning in Earthquake Seismology. Annual Review of Earth and Planetary Sciences, 51.

Moyer, P. A., Boettcher, M. S., McGuire, J. J., & Collins, J. A. (2018). Spatial and temporal variations in earthquake stress drop on Gofar Transform Fault, East Pacific Rise: Implications for fault strength. Journal of Geophysical Research: Solid Earth, 123(9), 7722-7740.

Novoselov, A., Sinkovics, K., & Bokelmann, G. This Earthquake Doesn't Exist. In NeurIPS 2021 AI for Science Workshop.

Park, Y., Mousavi, S. M., Zhu, W., Ellsworth, W. L., & Beroza, G. C. (2020). Machine-learning-based analysis of the Guy-Greenbrier, Arkansas earthquakes: A tale of two sequences. Geophysical Research Letters, 47(6), e2020GL087032.

Paszke A, Gross S, Massa F, et al. Pytorch: An imperative style, high-performance deep learning library[J]. Advances in neural information processing systems, 2019, 32.

Radford, A., Metz, L., & Chintala, S. (2015). Unsupervised representation learning with deep convolutional generative adversarial networks. arXiv preprint arXiv:1511.06434.

Ronneberger, O., Fischer, P., & Brox, T. (2015). U-net: Convolutional networks for biomedical image segmentation. In Medical Image Computing and Computer-Assisted Intervention–MICCAI 2015: 18th International Conference, Munich, Germany, October 5-9, 2015, Proceedings, Part III 18 (pp. 234-241). Springer International Publishing.

Ruthotto, L., & Haber, E. (2021). An introduction to deep generative modeling. GAMM-
43

Supplement to

**Deep generative model conditioned by phase picks for synthesizing labeled seismic waveforms with limited data**

By Guoyi Chen, Junlun Li* and Hao Guo

**Description of the Supplementary Material**

Supplementary materials for this paper includes synthetic waveforms generated by two alternative PhaseGen models, 10 randomly generated waveform samples by PhaseGen trained with batch normalization layers, change of $L_W$ with training epochs, statistics of maximum correlation coefficients for 1,000 randomly generated samples by PhaseGen trained with 20,000 epochs, trend in the mean of the maximum correlation coefficients with the training epochs, and 10 randomly generated waveform samples by PhaseGen directly using the P- and S-arrival times as two input conditions.

**Supplementary Figure Captions**

**Figure S1.** Generated waveforms by two alternative PhaseGen models trained with 200 different events indicated by the yellow dots in Figure 3a. These 200 events are divided equally into two groups for training the two models. (a) and (b) show 10 randomly generated waveforms by the two models, respectively. The red bars indicate the P-arrivals, and the blue bars indicate the S-arrivals.

**Figure S2.** Ten randomly generated waveforms by a PhaseGen model trained with batch normalization layers.

The red bars indicate the P-arrivals, and the blue bars indicate the S-arrivals.

**Figure S3.** Variation of $L_W$ over 20,000 epochs. The red star denotes the epoch adopted by the PhaseGen model in the main text.

**Figure S4.** Histogram illustrates the distribution of maximum correlation coefficients for 1000 randomly generated samples by an alternative PhaseGen model trained with 20,000 epochs. The mean and standard deviation of these coefficients are marked in the upper left corner.

**Figure S5.** Variation of the mean of the maximum correlation coefficients with the training epochs.

**Figure S6.** Ten randomly generated waveforms by PhaseGen directly using the P- and S-arrival as two input conditions. The red bars indicate the P-arrivals, and the blue bars indicate the S-arrivals.

**Supplemental Figures**

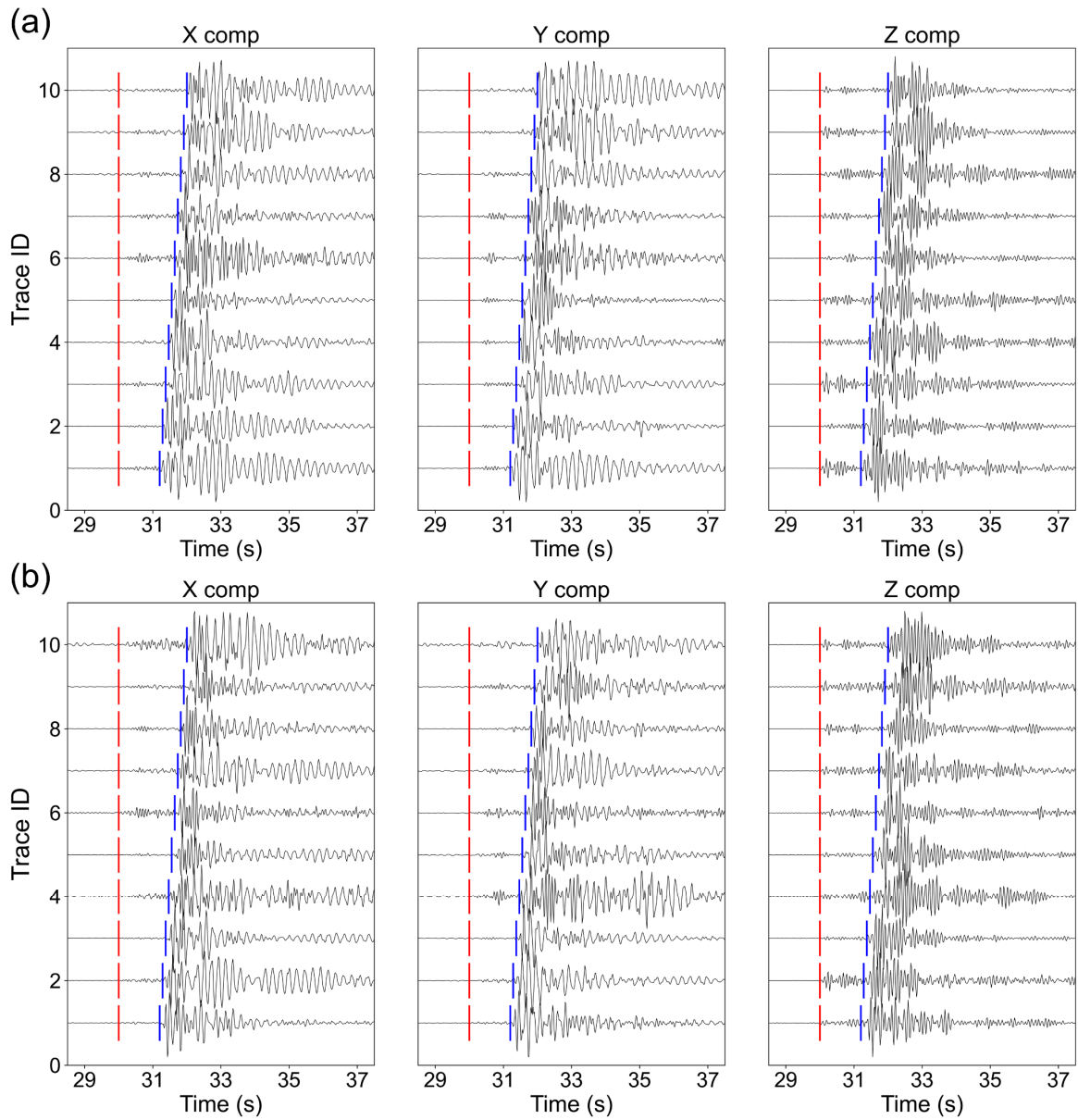

**Figure S1.**

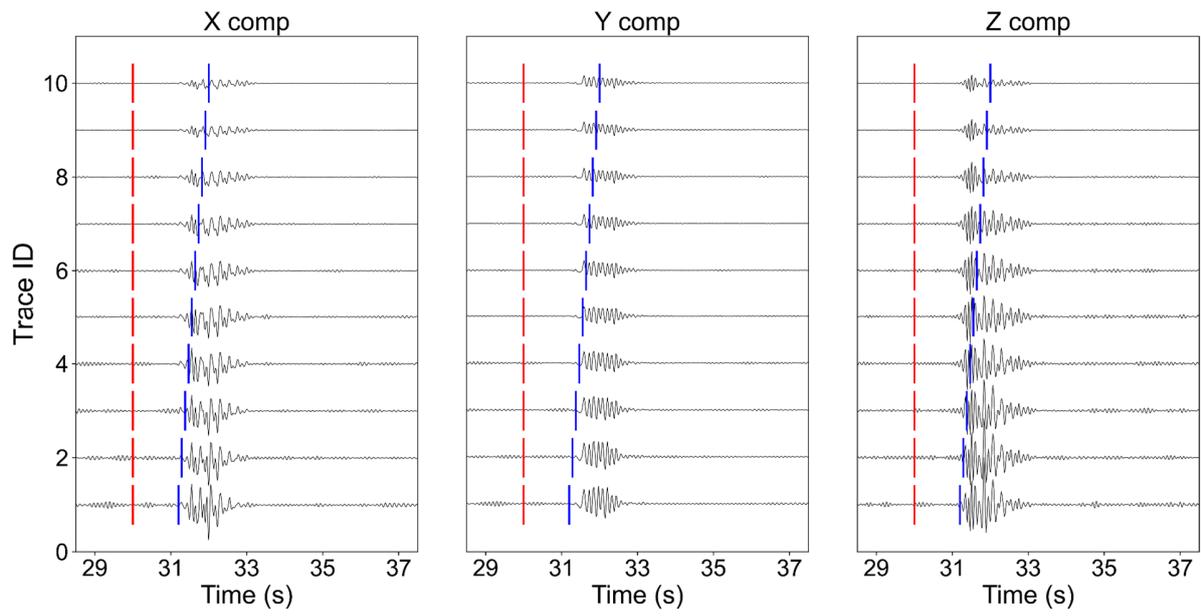

**Figure S2.**

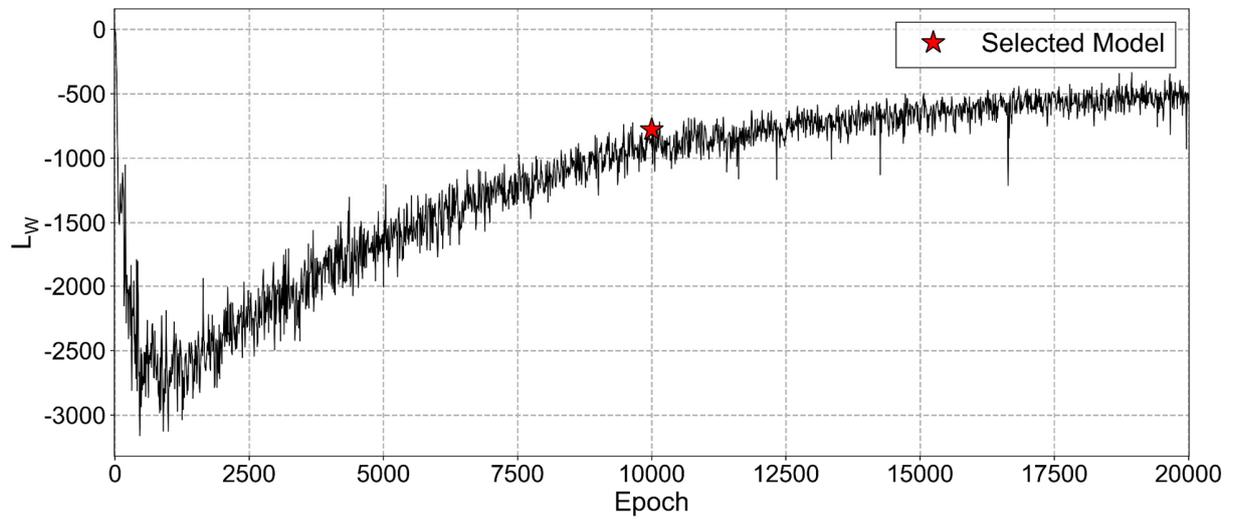

**Figure S3.**

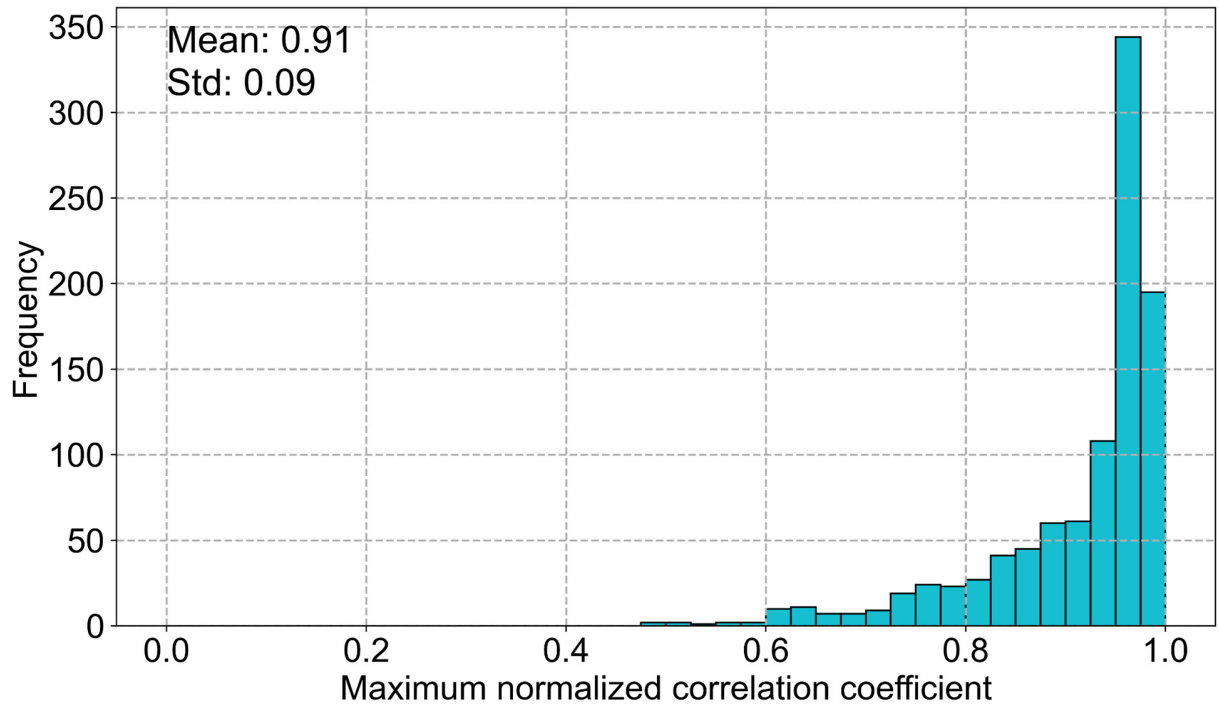

**Figure S4.**

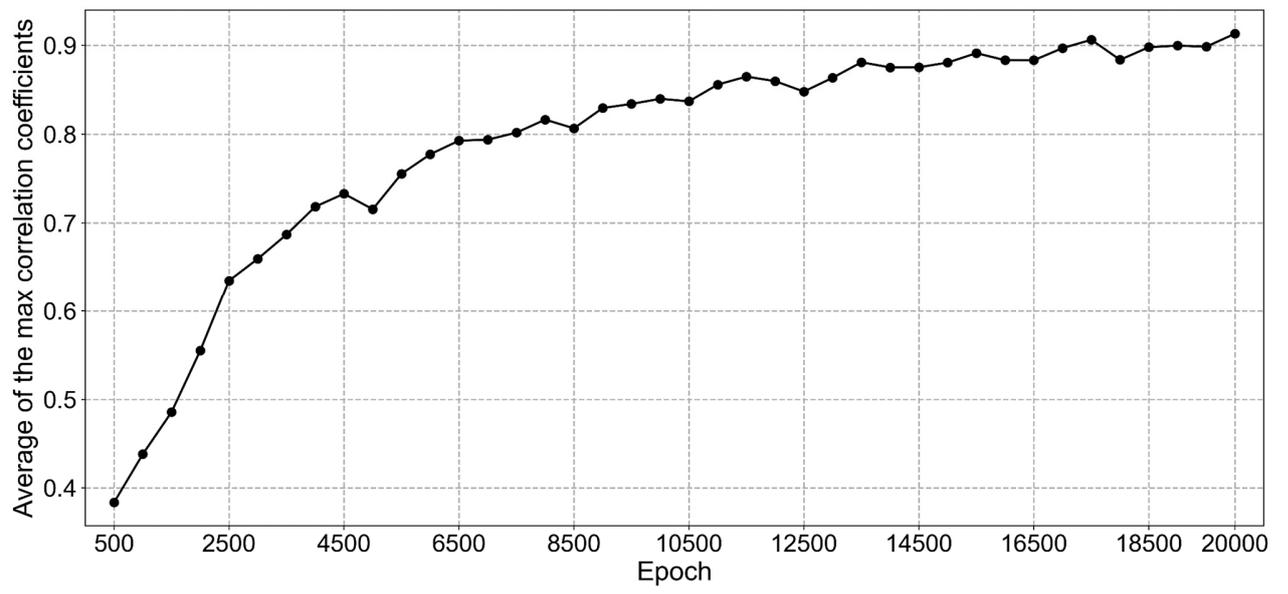

**Figure S5.**

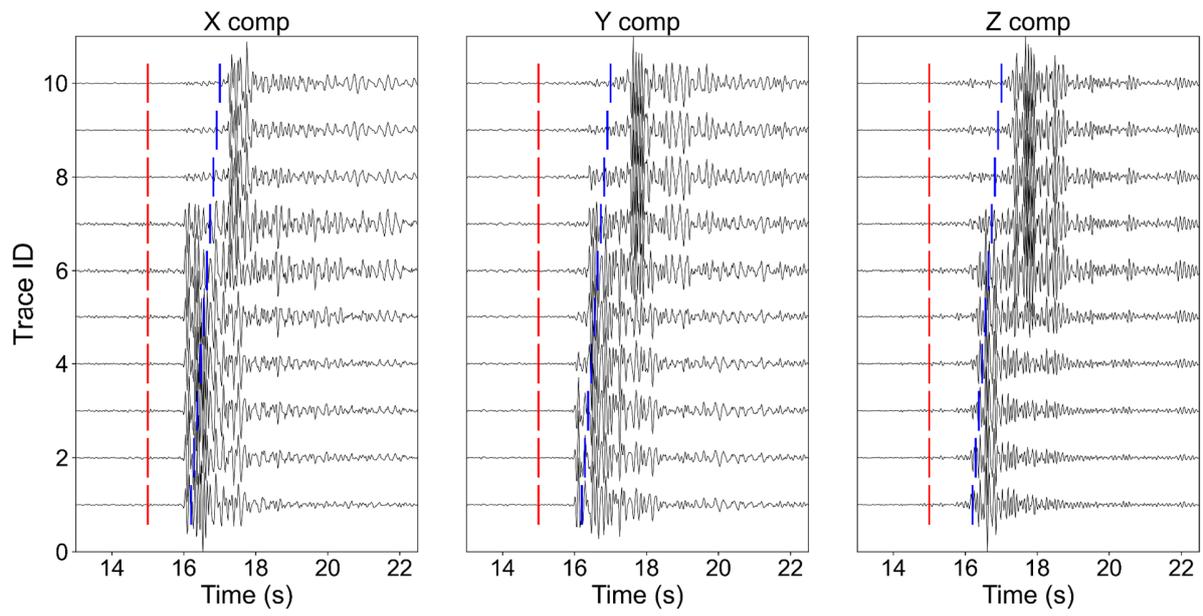

**Figure S6.**